\documentclass[iop]{emulateapj}
\slugcomment{}

\usepackage{natbib}
\usepackage{hyperref}

\begin{document}

\title{A Keplerian-like disk around the forming O-type star AFGL\,4176}
\shorttitle{A Keplerian-like disk around AFGL\,4176}
\shortauthors{Johnston et al.}

\author{Katharine G. Johnston\altaffilmark{1}, Thomas P. Robitaille\altaffilmark{2}, Henrik Beuther\altaffilmark{2}, Hendrik Linz\altaffilmark{2}, Paul Boley\altaffilmark{3}, Rolf Kuiper\altaffilmark{4,2}, Eric Keto\altaffilmark{5}, Melvin G. Hoare\altaffilmark{1} and Roy van Boekel\altaffilmark{2}}

\altaffiltext{1}{School of Physics \& Astronomy, E.C. Stoner Building, The University of Leeds, Leeds, LS2 9JT, UK; k.g.johnston@leeds.ac.uk}
\altaffiltext{2}{Max Planck Institute for Astronomy, K\"onigstuhl 17, D-69117 Heidelberg, Germany}
\altaffiltext{3}{Ural Federal University, Astronomical Observatory, 51 pr. Lenina, Ekaterinburg, Russia}
\altaffiltext{4}{Institute of Astronomy and Astrophysics, Eberhard Karls University T\"ubingen, Auf der Morgenstelle 10, D-72076 T\"ubingen, Germany}
\altaffiltext{5}{Harvard-Smithsonian Center for Astrophysics, 60 Garden St, Cambridge, MA 02138, USA}

\accepted{\today}

\newpage
\begin{abstract}
We present Atacama Large Millimeter/submillimeter Array (ALMA) line and continuum observations at 1.2\,mm with $\sim$0.3$''$ resolution that uncover a Keplerian-like disk around the forming O-type star AFGL\,4176. The continuum emission from the disk at 1.21\,mm (source mm1) has a deconvolved size of 870$\pm$110\,AU $\times$ 330$\pm$300\,AU and arises from a structure $\sim$8\,M$_{\odot}$ in mass, calculated assuming a dust temperature of 190\,K. The first-moment maps, pixel-to-pixel line modeling, assuming local thermodynamic equilibrium (LTE), and position-velocity diagrams of the CH$_3$CN J=13--12 K-line emission all show a velocity gradient along the major axis of the source, coupled with an increase in velocity at small radii, consistent with Keplerian-like rotation. The LTE line modeling shows that where CH$_3$CN J=13--12 is excited, the temperatures in the disk range from $\sim$70 to at least 300\,K and that the H$_2$ column density peaks at 2.8$\times$10$^{24}$\,cm$^{-2}$. In addition, we present Atacama Pathfinder Experiment (APEX) $^{12}$CO observations which show a large-scale outflow from AFGL\,4176 perpendicular to the major axis of mm1, supporting the disk interpretation. Finally, we present a radiative transfer model of a Keplerian disk surrounding an O7 star, with a disk mass and radius of 12\,M$_{\odot}$ and 2000\,AU, that reproduces the line and continuum data, further supporting our conclusion that our observations have uncovered a Keplerian disk around an O-type star.

\end{abstract}


\maketitle

\section{INTRODUCTION}

The process of star formation is often pictured as a young star being fed by a disk formed as a result of angular momentum conservation over a period of tens of thousands to millions of years. Yet this picture has been mainly derived from observations of disks that have practically finished accreting. In actuality, it has been confirmed only relatively recently that stable, rotationally supported, Keplerian disks around low-mass Class 0/I protostars exist at earlier times, when the star and its disk is concealed deeply in an accreting envelope \citep[e.g.,][]{brinch07b, jorgensen09, tobin12b, murillo13b}. 

The situation is even less clear when it comes to forming massive stars, which go on to reach a final mass of 8\,M$_{\odot}$ and above. As they form so quickly ($t_{\rm form} \sim$10$^{4}$ - 10$^{5}$ yr), they spend the entirety of their formation still embedded in a surrounding envelope. Are these protostars the high-mass equivalents of low-mass Class 0/I objects, and do they also have disks? Many of the energetic feedback mechanisms associated with massive stars, such as radiation pressure and at later times stellar winds and photo-ionization, would halt the vast amount of accretion required to form these stars. In theory, these can be bypassed by the existence of a disk \citep[e.g.][]{yorke020, krumholz09a, kuiper10, kuiper11}, allowing the radiation pressure, winds, or hot ionized gas to be channeled away along the axis perpendicular to the disk, where densities are lower.

Recent observations of early B-type (proto)stars have responded to this prediction with detections of disk candidates which have kinematics that appear to be dominated by the central protostar and are stable \citep[e.g.][]{johnston11,sanchez-monge13b,cesaroni14a,beltran14a}. Yet the number of disks discovered so far only constitutes a handful \citep{cesaroni07,beltran11a}, as many of the observed rotating structures, instead referred to as toroids, are too large and rotate too slowly to be in centrifugal equilibrium. Stepping up to O-type stars, there are even fewer candidates \citep[e.g.,][]{wang12,hunter14,zapata15b}; in these cases the detections are based on a velocity gradient across the source and an outflow or jet that is projected perpendicular to the candidate disk plane.

In this letter we present ALMA observations that trace a Keplerian-like disk toward the infrared source AFGL\,4176, which constitutes the best observational example of an O-type protostar with a disk to-date. 

AFGL\,4176 (G308.918+0.123, IRAS\,13395-6153) is a forming massive star with coordinates 13$^h$43$^m$01$^s.$69 -62$^{\circ}$08$'$51.3$''$ (FK5 J2000),  embedded in a star-forming region with a total luminosity of $\sim$10$^{5}$\,L$_{\odot}$ \citep[d = 4.2\,kpc,][]{boley12, green11a}. The source lies at the northern edge of an H\,\textsc{ii} region \citep[][]{caswell92,ellingsen05,shabala06} that peaks $\sim$4$''$ to the south of AFGL\,4176, likely powered by another star of spectral type O9. Four 6.7\,GHz Class II methanol maser spots with an extent of 840\,AU at 4.2\,kpc lie in close proximity to AFGL\,4176 along a line with position angle (PA) $\sim$-35$^{\circ}$ \citep{phillips98}. NH$_3$ observations have uncovered that the star is embedded in a large-scale rotating toroid with a radius of $\sim$0.7\,pc \citep{johnston14a}. The large-scale continuum emission at 1.2\,mm \citep{beltran06} traces a dense core of 0.8\,pc and 890\,M$_{\odot}$ at 4.2\,kpc, and knots of shocked H$_2$ emission have been detected \citep{de-buizer09}, suggesting the presence of an outflow. \citet{boley12} have modeled the spectral energy distribution (SED) and mid-IR interferometric observations of AFGL\,4176, finding that the latter required a non-spherically-symmetric model to adequately fit the data. They interpreted the mid-IR visibilities as a combination of a disk-like structure with radius of 660\,AU at 4.2\,kpc, inclination of 60$^{\circ}$, and PA=112$^{\circ}$, and a spherically symmetric Gaussian halo with a FWHM of $\sim$600\,AU \citep{boley12,boley13a}. Finally, \citet{ilee13a} found that the 2.3$\mu$m CO bandhead emission toward AFGL\,4176 is consistent with the inner $\sim$10\,AU of a Keplerian disk.

\section{OBSERVATIONS \label{observations}}

\subsection{ALMA Observations \label{ALMAobs}}
We observed AFGL\,4176 with the 12\,m antenna array of the Atacama Large Millimeter/submillimeter Array (ALMA) during Cycle 1, under program 2012.1.00469.S (PI Johnston). The observations were carried out on 2014 August 16 and 17 in dual-polarization mode in Band 6 ($\sim$250\,GHz or 1.2\,mm) under good weather conditions (precipitable water vapor, PWV$\sim$1.32 and 1.14\,mm respectively). AFGL\,4176 was observed with one pointing centered on 13$^h$43$^m$01$^s.$08 -62$^{\circ}$08$'$55.5$''$ (FK5 J2000). Two wide and two narrow spectral windows (spws) were observed with respective widths of 1.875\,GHz and 468.750\,MHz. The two wide spws were centered at 240.541 and 254.043\,GHz, while the two narrow spws were centered at frequencies of 239.072 and 256.349\,GHz. The spectral resolution was 1129\,kHz (1.41 and 1.33\,km\,s$^{-1}$) and 282\,kHz (0.354 and 0.330\,km\,s$^{-1}$) respectively.  Thirty-nine antennas were included in the array, of which 36 had useful data. Baseline lengths were 14.4 to 1210.8\,m, providing a largest angular scale of $\sim$18$''$. The primary beam size was 22.7 - 24.4$''$. The bandpass calibrators were J1617-5848 and J1427-4206 and absolute flux calibrators were Titan and Ceres on August 16 and 17 respectively. Phase/gain calibrators were J1308-6707 and J1329-5608 for both days. The flux calibration uncertainty was estimated to be $\lesssim$20\%. 

Calibration was carried out using the Common Astronomy Software Applications (CASA) version 4.2.1 via the delivered pipeline script. We improved the resulting images by self-calibration of the continuum, and applied these solutions to the line data. The continuum images were made using 1.4\,GHz bandwidth of line-free channels across all spws. The central frequency of the combined continuum emission is 247.689\,GHz (1.210\,mm). Imaging was carried out using Briggs weighting with a robust parameter of 0.5. The noise in the continuum image is 78\,$\mu$Jy\,beam$^{-1}$ in a beam of 0.28$''\times$0.24$''$, PA=-30.2$^{\circ}$. In this letter, we also present the detected K ladder transitions of CH$_3$CN J=13--12. The K=0 to 8 transitions were observed within the 239.072\,GHz band. For the K=2 to 8 images presented below, the noise ranges between 3.4 and 6.2\,mJy\,beam$^{-1}$ in a beam of 0.30$''\times$0.28$''$, PA=37.7 to 37.9$^{\circ}$. 

\subsection{APEX Observations \label{APEXobs}}
We observed AFGL\,4176 with the Atacama Pathfinder Experiment (APEX)\footnote{APEX is a collaboration between the Max-Planck-Institut f\"ur Radioastronomie, the European Southern Observatory, and the Onsala Space Observatory.} 12\,m antenna for program M0020\_89 during the night 2012 April 20--21 under very good weather (PWV$\sim$0.4\,mm). The Swedish Heterodyne Facility Instrument (SHeFi) APEX-2 receiver was tuned to $^{12}$CO(3-2) at 345.79599\,GHz in the lower sideband, providing a beam size of 18$''$. On-the-fly maps of 2.5$' \times$2.5$'$ extent were taken in two perpendicular scan directions to reduce scanning artifacts. The XFFTS2 backend provided a channel separation of 0.1984\,km\,s$^{-1}$. The achieved rms in the spectra in the central part of the map is around 0.12 K. The data reduction was performed within GILDAS/CLASS\footnote{http://www.iram.fr/IRAMFR/GILDAS}. Brightness temperatures throughout this letter are stated in main-beam temperatures (T$_{\rm MB}$). 

\section{RESULTS and DISCUSSION \label{results}}

\begin{figure}
\epsscale{1.2}
\plotone{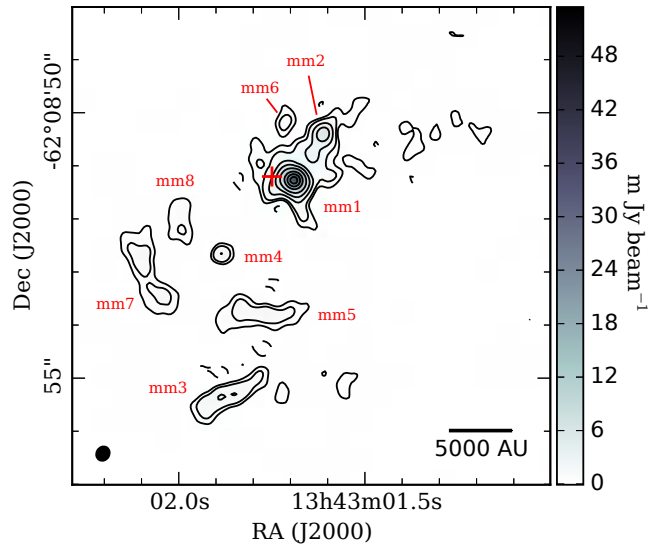}
\caption{Continuum emission toward AFGL\,4176 at 1.21\,mm observed with ALMA, in grayscale and contours ($\sigma$ = 78$\mu$Jy\,beam$^{-1} \times$ -5, 5, 10, 25, 50, 100, 200, 300, 400). The beam is shown in the bottom left corner. The red cross shows the position of the Class II methanol maser group reported in \citet{phillips98} and the mm sources are labeled. \label{contfig}}
\end{figure}

\begin{figure}
\epsscale{1.2}
\plotone{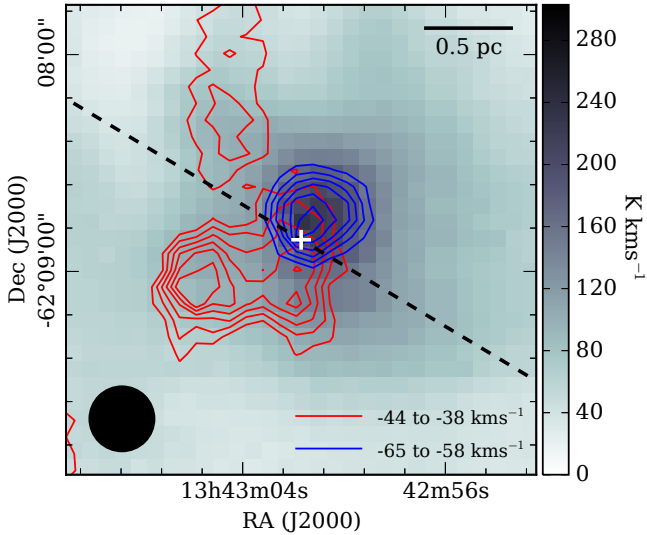}
\caption{Large-scale bipolar outflow from AFGL\,4176 seen in $^{12}$CO J=3--2 with APEX. The total integrated emission is shown in grayscale, and the integrated red- and blue-shifted emission is shown in contours at 40, 50, 60...90\,\% of the peak integrated fluxes (11.5 and 24.9\,K km\,s$^{-1}$ respectively). Velocity ranges for integration of the red- and blue-shifted emission are shown in the bottom right, and the beam in the bottom left. The cross and dashed line mark the peak and PA of the continuum emission shown in Fig.\,\ref{contfig}. \label{12COfig}}
\end{figure}

\begin{figure*}
\epsscale{1.1}
\plotone{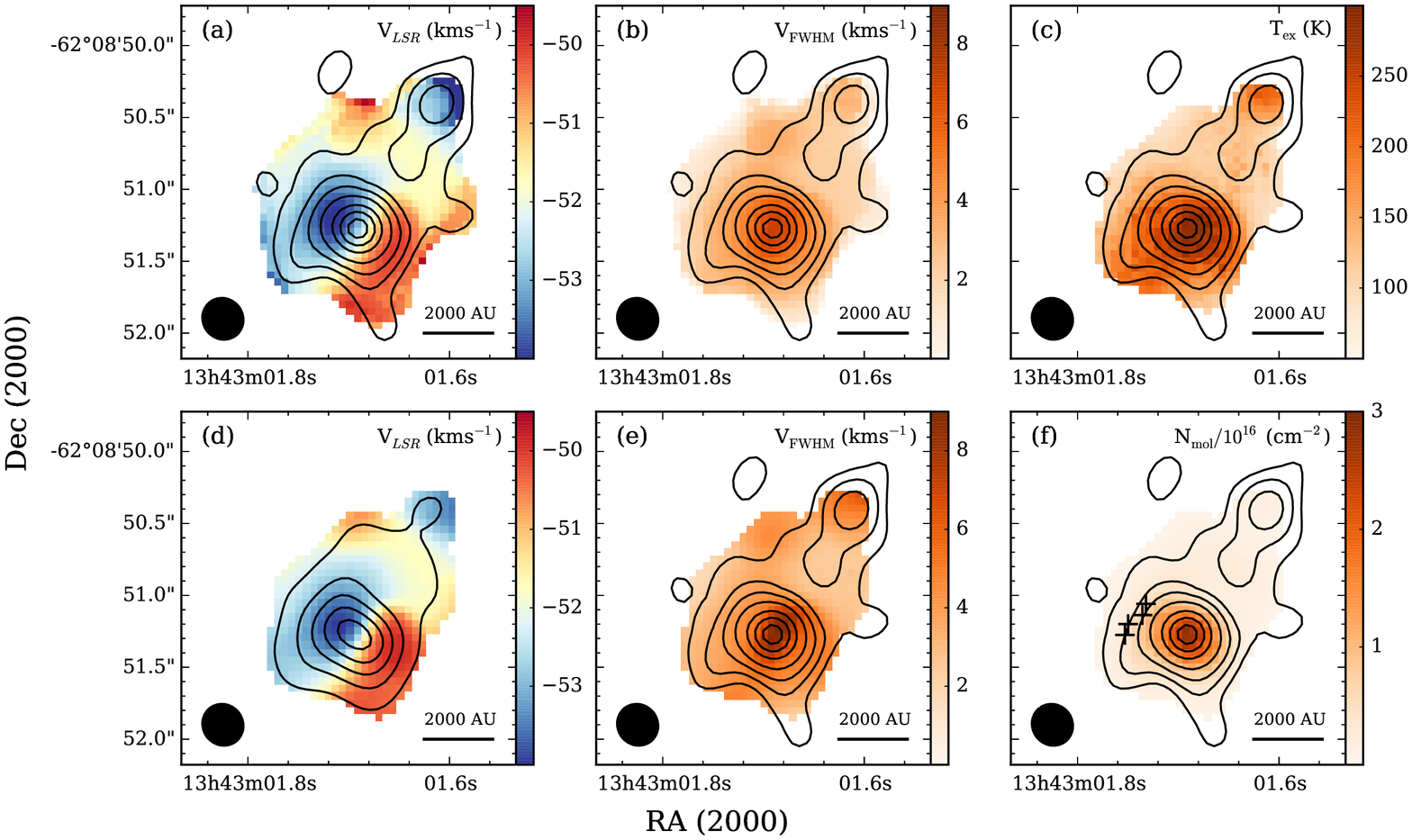}
\caption{Panels a) and b) show first- and second-moment maps of the CH$_3$CN J=13--12, K=3 emission from AFGL\,4176 mm1 in colorscale. Panels c) through f) show the results of the CASSIS pixel-to-pixel spectrum fitting. Continuum emission (starting at 10\,$\sigma$) similar to that shown in Fig.\,\ref{contfig} is overplotted as contours in all panels except d), which shows the integrated K=3 emission at 10, 30, 50, 70 and 90\% of the peak. The beam is shown in the bottom left corner of all panels. Panel f) shows the positions of the Class II methanol masers from \citet{phillips98}. \label{CH3CNfig}}
\end{figure*}

In Fig.\,\ref{contfig} we present the 1.21\,mm continuum emission from the AFGL\,4176 region. The emission is dominated by mm1, with a peak position in the non-self-calibrated image of 13$^h$43$^m$01$^s.$693 -62$^{\circ}$08$'$51.25$''$ (FK5 J2000), coincident with the position of AFGL\,4176 in 2MASS (with $\sim$0.1$''$ positional uncertainty\footnote{http://www.ipac.caltech.edu/2mass/releases/allsky/doc/ sec1\_6b.html\#psrecptsour}). Performing a gaussian fit to mm1, we determined a peak flux of 37$\pm$2\,mJy\,beam$^{-1}$, an integrated flux of 50$\pm$4\,mJy, a deconvolved size of 0.21$\pm$0.03$'' \times$0.08$\pm$0.07$''$ (870$\pm$110\,AU $\times$330$\pm$300\,AU), and PA=59$\pm$17$^{\circ}$. Using the equations of \citet{hildebrand83}, as well as assuming 190\,K (the average temperature of the disk derived from CASSIS LTE line modeling below), a gas-to-dust ratio of 154 \citep{draine11}, and a dust opacity at 1.21\,mm of 0.24\,cm$^2$\,g$^{-1}$ \citep[][with $R_V=5.5$]{draine03a,draine03b}, we derive a gas mass and peak column density of 8\,M$_{\odot}$ and 8$\times$10$^{24}$\,cm$^{-2}$ for mm1. As well as a compact component, mm1 includes low-lying 5-10$\sigma$ emission extending NW, perpendicular to the major axis of mm1, that joins it to mm2. The remaining mm sources lie in the NW and SE quadrants around mm1, including two compact sources mm2 and mm4 on opposite sides of mm1. Full observed properties of all detected 1.21\,mm sources will be given in a upcoming paper (Johnston et al., in preparation).

Figure\,\ref{12COfig} shows the integrated $^{12}$CO J=3--2 emission from the red- and blue-shifted high-velocity wings of the bipolar outflow from AFGL\,4176 observed by APEX, overlaid upon a $^{12}$CO zero-moment map shown in grayscale. The outflow lobes are orientated roughly NW-SE, perpendicular to the PA of the source mm1, shown as a dashed line in Fig.\,\ref{12COfig}. This relative geometry suggests mm1 is a disk driving the large-scale outflow~seen~in~$^{12}$CO.

The kinematics of the gas in mm1 traced by first- and second-moment maps (intensity-weighted velocity and linewidth fields respectively) of CH$_3$CN J=13--12 K=3 emission are shown in panels a) and b) of Fig.\,\ref{CH3CNfig}. The velocity field of the K=3 line in panel a) shows a clear velocity gradient along the major axis of mm1, which is also present in lines K=2--8 and on arcminute scales in NH$_3$ \citep{johnston14a}. The velocity field shown in Fig.\,\ref{CH3CNfig} is similar to that expected from disks in near-Keplerian rotation \citep[e.g. HD\,100546 and TW Hya, respectively][]{pineda14a, hughes11a}. For instance, there is a quick change from blue- to red-shifted emission when crossing the minor axis of the source. In addition, the emission in the most blue- and red-shifted channels is found close to the continuum peak position, whereas the lower-velocity gas is more extended. The linewidths across mm1 in panel b) peak towards the continuum peak, consistent with beam-averaging of high-velocity red- and blue-shifted emission. There is extended CH$_3$CN emission to the NW of mm1 with near-systemic velocities (v$_{_{\rm LSR}}\sim$-52\,km\,s$^{-1}$), as well as blue-shifted gas associated with mm2, which would be consistent with this source being associated with the blue-shifted lobe seen in $^{12}$CO.

Figure\,\ref{CH3CNfig} panels c) through f) present the results of modeling the CH$_3$CN and CH$_3^{13}$CN J=13--12 K-ladders in the spectrum associated with each pixel, making a map of excitation temperature ($T_{\rm ex}$), velocity (v$_{_{\rm LSR}}$), linewidth (v$_{_{\rm FWHM}}$), and column density ($N_{\rm mol}$) across the source. The modeling was carried out using CASSIS\footnote{CASSIS is developed by IRAP-UPS/CNRS (http://cassis.irap.omp.eu)} and the JPL molecular spectroscopy database\footnote{http://spec.jpl.nasa.gov}, using the Markov Chain Monte Carlo $\chi^2$ minimization method and assuming Local Thermodynamic Equilibrium (LTE). Given $N_{\rm mol}$ and $T_{\rm ex}$, CASSIS determines an optical depth which is included in each model. We assume a [$^{12}$C/$^{13}$C] abundance ratio of 60 \citep[assuming the distance to the Galactic center is 8.4\,kpc and thus a Galactocentric distance to AFGL\,4176 of 6.6\,kpc;][]{reid09, milam05a}, a source size of 0.53$''$ from the fitted size of the K=0,1 zero-moment map, and initial parameter ranges $T_{\rm ex}$ = 50 -- 350\,K, v$_{_{\rm LSR}}$ = -57 -- -47\,km\,s$^{-1}$, v$_{_{\rm FWHM}}$ = 0.5 -- 10\,km\,s$^{-1}$ and $N_{\rm mol}$ = 1$\times$10$^{14}$ -- 1$\times$10$^{17}$\,cm$^{-2}$. The resultant temperature shown in panel c) peaks at the continuum peak position, and has values ranging between 74 and 294\,K. There is also a secondary temperature peak toward mm2. Panels d) and e) show similar results to the first- and second-moment maps shown in panels a) and b), however an increase in linewidth can be seen more clearly toward mm2. The column density of CH$_3$CN peaks at 2.8$\times$10$^{16}$\,cm$^{-2}$ toward the center of mm1, with a slight elongation along the major axis. Using the assumed size above, and an abundance of CH$_3$CN of 10$^{-8}$ \citep[e.g.][]{gibb00a, bisschop07a}, this corresponds to H$_2$ column and volume densities of 2.8$\times$10$^{24}$\,cm$^{-2}$ and $>$8$\times$10$^{7}$\,cm$^{-3}$ respectively. 

Having established that a rotating structure, likely a disk, surrounds AFGL\,4176 mm1, we can compare it with the geometry of the circumstellar material determined from modeling of observations from the MID-infrared interferometric Instrument \citep[MIDI;][]{boley12,boley13a}. The PA of mm1 ($\sim$60$^{\circ}$) is not aligned with the 112$^{\circ}$ determined from the MIDI visibilities. Although these PA are not exactly orthogonal, these two observations may be reconciled if the MIDI observations are instead tracing heated dust in the outflow cavity, such as found for W33\,A \citep{de-wit10a}. We also note that although offset from the continuum peak of mm1, the four Class II methanol maser spots detected by \citet{phillips98} shown in Figure\,\ref{CH3CNfig} f) lie along a line almost perpendicular to the disk.

\begin{figure}
\epsscale{1.2}
\plotone{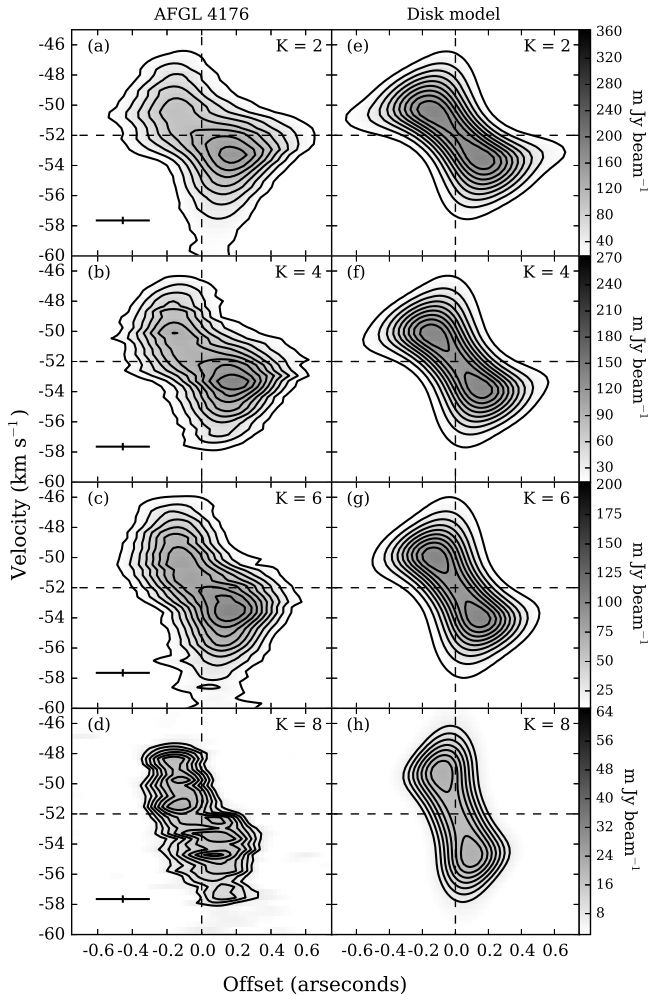}
\caption{Panels a) through d): position-velocity diagrams of CH$_3$CN J=13--12 K=2,4,6, and 8 averaged along a cut centered on the mm1 continuum peak position, with PA=61.5$^{\circ}$ and width=1$''$. The horizontal and vertical dashed lines mark the position of the continuum peak and a velocity of -52\,km\,s$^{-1}$. The crosses in the bottom left of each panel show the observational spatial and spectral resolution. Panels e) through h): position-velocity diagrams of the same lines for the model described in the text. Contour levels are 10,20,30...90\% of the peak flux, except for K=8 which instead starts at 30\%. \label{PVfig}} 
\end{figure}

To investigate the dynamics of mm1 further, we present position-velocity (PV) diagrams of the CH$_3$CN J=13--12 K=2,4,6, and 8 lines in Fig.\,\ref{PVfig} panels a) to d), along a cut centered on the continuum peak position of mm1, with width=1$''$ and PA=61.5$^{\circ}$, which was the average PA of the emission in the CH$_3$CN J=13--12 K=4, 5 and 6 lines, which had good signal-to-noise and were not contaminated by envelope emission.

The PV diagrams of the K=2,4 and 6 lines exhibit the ``butterfly'' shape expected from a Keplerian-like rotation curve that increases in velocity close to the central object, and is more extended close to the systemic velocity. It  bears close resemblance to the PV-diagrams observed toward the forming early B-type star IRAS\,20126+4014, which has a disk in Keplerian-like rotation \citep{cesaroni05a, cesaroni14a}, as well as the forming O-type stars NGC 6334 I(N) SMA\,1b and IRAS\,16547-4247 \citep[][respectively]{hunter14, zapata15b}. The velocity gradient of the K-lines becomes steeper at higher values of K, thus the more-excited lines that trace hotter gas closer to the central object also trace higher velocities, which is expected from Keplerian-like rotation.

Interestingly, the PV-diagrams of the lower-K lines shown in Fig.\,\ref{PVfig} are asymmetrical about the systemic velocity, with the blue-shifted emission being brighter. This is not observed however in the K=7,8 lines. A similar blue-asymmetry is seen in the PV-diagrams of the low-excitation K-lines of CH$_3$CN J=12--11 observed by \citet{cesaroni14a} and \citet{hunter14}, whereas the high-excitation lines are either symmetrical or exhibit a red-asymmetry. The CH$_3$CN J=19--18 K=2 PV diagram of G35.03+0.35\,HMC\,A also shows this blue-asymmetry \citep{beltran14a}.  This has been suggested to be due to an asymmetric disk structure, however the fact that so many objects exhibit the same features in their PV-diagrams suggests this is more likely due to a radiative transfer effect and/or a geometry that is present in all sources.

To check whether a Keplerian-disk model is consistent with our observations, we ran a grid of self-consistent gas and dust radiative transfer models, where the line and continuum radiative transfer were performed using the codes \textsc{Mollie} \citep[assuming LTE,][]{keto10a} and \textsc{Hyperion} \citep{robitaille11}, using the Milky Way dust properties from \citet{draine03a,draine03b} with $R_V=5.5$, respectively. To fit the models to the line and continuum observations, we fit the profiles of the continuum and CH$_3$CN J=13--12 K=2-8 emission collapsed along the major and minor axes, as well as the integrated spectra for the lines. The models were convolved to the observed beam before fitting. 

Panels e) to h) of Fig.\,\ref{PVfig} show the CH$_3$CN J=13--12 PV diagrams for a model which provides a good fit to the line and continuum data. The model consists of a Keplerian flared disk of radius 2000\,AU (the inner radius is set to be the dust sublimation radius, in this case 31.3\,AU), total gas mass 12\,M$_{\odot}$, with a surface density decreasing as $r^{-1.5}$ and an inclination of 30$^{\circ}$. The scaleheight of the disk is given by $z=6.7~(\varpi/100 \rm{AU})^{1.29}$\,AU where $\varpi$ is the cylindrical radius and the constants were determined in order for the disk to be in hydrostatic equilibrium. The model includes a rotationally-flattened infalling envelope \citep{ulrich76} with an infall rate of 4.6$\times10^{-4}$\,M$_{\odot}$\,yr$^{-1}$ and an outer radius of 150,000\,AU. The central object is a 25\,M$_{\odot}$ zero-age main-sequence O7 star, with properties determined from \citet{meynet00a} and \citet{martins05}, specifically T=36,872\,K and R=6.71\,R$_{\odot}$. The Keplerian velocity field in the model accounts for the circumstellar mass interior to a given radius. The abundance of CH$_3$CN relative to H$_2$ is taken to be 10$^{-8}$ at $>$100\,K, dropping to 5$\times$10$^{-9}$ for 90\,K $<$ T $<$ 100\,K, and again to 10$^{-10}$ for $<$90\,K \citep{gerner14a, collings04a}. The grid of models as well as more detailed results will be presented in a future publication (Johnston et al., in preparation). 

We note that the model disk mass of 12\,M$_{\odot}$ differs from that derived using the single temperature of 190\,K (8\,M$_{\odot}$) because it accounts for the expected variance in temperature and optical depth over the disk, and is the best fit from a discrete set of disk masses. In addition, a Gaussian fit to the continuum model image gives a major FWHM of 880\,AU, in good agreement with the fit to the observations (FWHM=870\,AU), which indicates the Gaussian fit underestimates the actual disk size.

\section{CONCLUSIONS \label{results}}

We present ALMA observations which uncover the presence of a Keplerian-like disk around an O-type star, detected in both 1.21\,mm continuum and CH$_3$CN. The velocity structure as traced by CH$_3$CN moment maps and LTE line modeling shows a clear velocity gradient and evidence of higher velocities at small radii, as expected for a Keplerian disk. In addition, we present a radiative transfer model which agrees with our ALMA observations not only in terms of the morphology of the PV diagram, but also in terms of the absolute flux of the lines and the continuum, and corroborates that we have uncovered the best example to-date of a disk in Keplerian-like rotation around a forming O-type star.

\acknowledgments

We thank the referee for insightful comments that helped improve this letter. We are grateful to have been able to observe with ALMA and APEX on Llano de Chajnantor, Chile. ALMA is a partnership of ESO (representing its member states), NSF (USA) and NINS (Japan), together with NRC (Canada), NSC and ASIAA (Taiwan), and KASI (Republic of Korea), in cooperation with the Republic of Chile. The Joint ALMA Observatory is operated by ESO, AUI/NRAO and NAOJ. We thank Frank Wyrowski, Miguel Angel Requena Torres, and our ALMA contact scientist Edwige Chapillon. PB acknowledges support from the Russian Science Foundation, grant No. 15-12-10017.

\end{document}